\documentclass{article}
\usepackage{amssymb}

\input{tcilatex}
\begin{document}

\title{Matter-antimatter Asymmetry, CP Violation and the Time Operator in
Relativistic Quantum Mechanics}
\author{M. Bauer \\
Instituto de F\'{\i}sica, Universidad Nacional Aut\'{o}noma de M\'{e}xico,\\
Ciudad Universitaria, CP 04510, M\'{e}xico, CDMX, MEXICO\\
e-mail: bauer@fisica.unam.mx}
\maketitle

\begin{abstract}
The matter antimatter asymmetry in the Universe is one of the outstanding
problems in physics.and cosmology. CP violation (CPV) is a necessary
condition for generating such asymmetry but the amount predicted in the
Standard Model (SM) from quark oscillations is too small by several orders
of magnitude. The dynamical time operator derived together with Dirac's
Hamiltonian from the Dirac canonical quantization of special relativity is
shown to exhibit a CPV inversely dependent on mass. Applying to all
fermions, it follows that extended Standard Models (ESM) with massive
neutrinos contains a much larger CPV as the electron and muon neutrino
masses are six to nine orders smaller than the up and down quarks masses.
Furthermore, it also follows that the CPV differs between neutrino and
antineutrino, in agreement with the T2K observations.
\end{abstract}

\bigskip

\bigskip

The matter antimatter asymmetry in the Universe is one of the outstanding
problems in physics.and cosmology. Following Sakharov\cite{Sakharov}, CP
violation (CPV) is a necessary condition for generating such asymmetry \ In
the Standard Model (SM) quark oscillations introduce a CPV phase but it is
found too small by several orders of magnitude $(\sim 10^{-10})$ to account
for the asymmetry observed\cite{Gavela,Huet}. The extended Standard Model
(ESM) that incorporates massive neutrinos does provide an additional CPV.
This is currently the experimental subject of the T2K collaboration where
the muon neutrino to electron neutrino oscillations and the corresponding
antineutrino oscillations are measured. They are already found to be
different, indicating a possibly large CPV as the hypothesis of CP
conservation is excluded at 90\% confidence level\cite{Nature,CERN,Miller}.
The data also indicate that muon neutrinos transform into electron neutrinos
more readily than muon antineutrinos transform into electron antineutrinos.\
Finally they seem to favour the normal ordering of the neutrino masses.

In the present letter it is shown that the existence of an internal time
property, derived from the canonical quantization of Special Relstivity (SR)
and common to all fermions, already exhibits a much larger CPV in an
extended Standard Model (ESM) with oscillating massive neutrinos than the
one derived from the Standard Model (SM) that considers only quark
oscillations. This follows from the inverse relation between this CPV and
the fermion mass.

Dirac canonical quantization, together with Born's Reciprocity Principle\cite%
{Born,Born2,Freidel}, of the scalar invariants of special relativity\cite%
{Aguillon}, namely $p^{\mu }p_{\mu }$, $x^{\mu }x_{\mu }$ and $\{x^{\mu
}p_{\mu }-p^{\mu }x_{\mu }\}$, results in quantum mechanical constraints
that yield both the Dirac Hamiltonian $H_{D}=c\mathbf{\alpha .p}+\beta
m_{0}c^{2}$ and an associated time operator $T=\mathbf{\alpha .r}/c+\beta
\tau _{0\text{,}}$ where $\tau _{0}$\ is a time invariant introduced in
analogy to the rest mass\footnote{%
The derivation requires the following conmutation relations $\left[ \hat{x}%
_{\mu },\hat{x}_{\nu }\right] =\left[ \hat{p}_{\mu },\hat{p}_{\nu }\right]
=\delta _{\mu \nu }$ and $\left[ \hat{x}_{\mu },\hat{p}_{\nu }\right]
=i\hbar \delta _{\mu \nu }$ be satisfied and the $\alpha _{i}$\ \ and $\beta 
$\ \ matrices obey a Clifford algebra.}. The equations satisfied by these
operators are shown to be:

a) the well known eigenvalue Dirac equation

\begin{equation}
\{c\mathbf{\alpha .\hat{p}}+\beta m_{0}c^{2}\}\left\vert \Psi \right\rangle
=e\ \left\vert \Psi \right\rangle \text{ \ with}\ \{cp_{0}:=e\}
\end{equation}%
that yields a continuum energy spectrum from \ $-\infty $\ to $+\infty $\ ,
except for a gap $2m_{0}c^{2}$\ at \ the origin; and similarly

b) an eigenvalue equation for an internal time variable $\tau $:%
\begin{equation}
\{\mathbf{\alpha .\hat{x}}/c+\beta \tau _{0}\}\left\vert \Psi \right\rangle
=\tau \left\vert \Psi \right\rangle \text{ \ with\ }\{x_{0}/c:=\tau \}
\end{equation}%
with a continuum time spectrum from $-\infty $\ to $+\infty $\ , except for
a gap $2\tau _{0}$\ at \ the origin. It is furthermore shown that \ $\tau
_{0}$ is equal to $h/m_{0}c^{2}$ (the deBroglie period)\cite{Bauer3}, making
the gaps complementary: to a large energy gap corresponds a small time gap,
and viceversa. The existence of this internal time operator for fermions
restores in RQM the equivalent footing accorded to space and time in SR,
circumventing Pauli's objection\cite{Pauli} to the existence of a time
operator\ dependent on the dynamical variables, that is accepted explictly
or implicitly in most of the extensive literature on the problem of time in
quantum mechanics\cite{Muga,Muga2}.

Denoting as $<>_{P},<>_{C},<>_{T}$ the expectation values in parity
inverted, charge conjugated, time reversed states respectively, one has\cite%
{Greiner,Aguillon}:%
\[
<\boldsymbol{r}>_{P}=-<\boldsymbol{r}>;<\boldsymbol{p}>_{P}=-<\boldsymbol{p}%
>;<\boldsymbol{\alpha }>_{P}=-<\boldsymbol{\alpha }>;<\beta >_{P}=-<\beta >
\]

\[
<\boldsymbol{r}>_{C}=<\boldsymbol{r}>;<\boldsymbol{p}>_{C}=-<\boldsymbol{p}%
>;<\boldsymbol{\alpha }>_{C}=<\boldsymbol{\alpha }>;<\beta >_{C}=-<\beta >
\]

\[
<\mathbf{r}>_{T}=<\mathbf{r}>;<\mathbf{p}>_{T}=-<\mathbf{p}>;<\mathbf{\alpha 
}>_{T}=-<\mathbf{\alpha }>;<\beta >_{T}=<\beta >
\]%
Then, from its definition, it follows that the time operator is not
invariant under the CP transformation. Indeed:%
\begin{equation}
\left\langle T\right\rangle _{CP}-\left\langle T\right\rangle
=-2\left\langle \beta \right\rangle \tau _{0}=-2\left\langle \beta
\right\rangle h/m_{0}c^{2}
\end{equation}%
where the subscript indicates expectation value in the CP transformed state.
As the neutrino masses, still not fully determined but expected to be of the
order of $10^{-3}$ to $10^{-1}$ev, and \ thus $10^{-9}$ to $10^{-6}$\ of the
masses of the up and down quarks, their associated CPV is$\ 6$ to $9$ orders
of magnitude larger than the CPV associated with the quark oscillations. \
Furthermore, the normal ordering of the neutrino masses is needed to insure
that it is the electron and muon neutrinos that provide the maximum CPV.

Another consequence arises from the dependence on $<\beta >$ that
distinguishes wavepackets of positive times $<\beta >\simeq 1$ (associated
with particles) and wavepackets of negative times $<\beta >\simeq -1$
(associated with antiparticles). The CPV for paticles is negative, so their
internal time suffers a delay in the CP state, while for antiparticles the
CPV it is positive and their internal time is advanced in the CP state. The
associated laboratory times in the Time Dependent Schr\"{o}dinger Equation
(TDSE), as introduced below, are converesily advanced for a particle respect
to its.antiparticle.

Indeed, the time parameter $t$ in the TDSE of an atomic system has been
shown to derive from its entanglement with the macroscopic classical
environment which then monitors its time development. It corresponds to the
time of the laboratory clock\cite{Briggs3,Briggs}, thus yielding a two times
perspective to the problem of time in RQM.

Considering that%
\begin{equation}
\lbrack \hat{T},\hat{H}_{D}]=i\hbar \{1+2\beta K+2\beta (\tau _{0}\hat{H}%
_{D}-m_{0}c^{2}\hat{T})\}
\end{equation}%
where $K=\beta (2\mathbf{s.l}/\hbar ^{2}+1)$ \ is a constant of motion\cite%
{Thaller},\ the Heisenberg picture equation satisfied by $\hat{T}(t)$ is%
\begin{equation}
\frac{d\hat{T}(t)}{dt}=\{1+2\beta K\}+(2\beta /i\hbar )\{\tau _{0}\hat{H}%
_{D}-m_{0}c^{2}\hat{T}(t)\}
\end{equation}%
Integrating from $0$ to $t:$%
\begin{equation}
\hat{T}(t)=\hat{T}(0)+\mathbf{\alpha (}0\mathbf{).}\{c\mathbf{\hat{p}/}%
H_{D}\}t+[(\exp -2iHt/\hbar )-1][(\hbar c/2iH)\mathbf{F/}c+G\tau _{0}]
\end{equation}%
where $G=\beta (0)-m_{0}c^{2}/H$.and.$\mathbf{F/}c=\{\mathbf{\alpha (}0%
\mathbf{)}-c\mathbf{p}/H\},$.whereas similarly\cite{Thaller},:%
\begin{equation}
\mathbf{\hat{r}(}t\mathbf{)=\hat{r}(}0\mathbf{)+}\{c^{2}\mathbf{\hat{p}/}%
H_{D}\}t+[(\exp -2iHt/\hbar )-1]\frac{i\hbar c}{2H}\mathbf{F/}c
\end{equation}%
Therefore $\hat{T}(t)$ exhibits a lineat $t$-dependence superimposed by an
oscillation (Zitterbewegung) of frequency $<2H/\hbar >,$ in a similar way to 
$\mathbf{\hat{r}(}t\mathbf{)}$. However as $\hat{T}(0)=\mathbf{\alpha (}0%
\mathbf{).r(}0\mathbf{)}/c+\beta (0)\tau _{0\text{,}}$ it mantains the
dependence on $\beta $ noted above, the evolution of $\hat{T}(t)$\ differs
in starting point and linear dependence on $t$ for $\left\langle \beta
(0)\right\rangle \simeq 1$ and \ $\left\langle \beta (0)\right\rangle \simeq
-1$\ \ wave packets in the laboratory. As, from Eq.6, the laboratory
evolution time is inversely propotional to the internal time, the neutrinos
will advance more rapidly than the antineutrinos and attain the detection
position earlier. Or equivalently, at a certain time t at the location of
the detector, more neutrinos than antineutrinos are expected, accounting for
the increased rate of oscillations from muon to electron neutrinos obseved.

Both the dependence on mass of an internal time and the monitoring of the
atomic system by its entangled macroscopic environment have already received
experimental confirmation\cite{Lan,Moreva,Moreva2}

In conclusion a CPV associated to all fermions results from the\ canonical
cuantization of SR as it reveals the existence of an internal time
observable \ Furthemore the CPV is inversely proprtional to the fermion
mass, establishing allready a six to nine of magnitude difference between
electron and muon neutrinos and the lightest quarks. As such it may be
sufficient to account for the matter antimatter asymmetry observed. It
supports the normal ordering of the neutrino masses. In addition it
exhibits.a difference between the neutrinos and antineutrinos evolutions in
the laboratory that may be in accordance with T2K observations, where the
data suggest that muon neutrinos transform into electron neutrinos more
readily than muon antineutrinos transform into electron antineutrinos\cite%
{Nature}. Alltogether it represents another important consequence of Special
Relativity.

\end{document}